# Martian seismic anisotropy underneath Elysium Planitia revealed by direct *S* wave splitting

Jing Shi[1], Cunrui Han[2], Tao Wang[1*], Chao Qi[3,4*], Han Chen[1], Zhihan Yu[1], Jiaqi Geng[1], Minghan Yang[1,5], Xu Wang[6], Ling Chen[4,6], Hejiu Hui[1,7]

[1]Frontiers Science Center for Critical Earth Material Cycling, School of Earth Sciences and Engineering, Nanjing University, Nanjing, China.

[2]School of Natural Sciences, Birkbeck, University of London, London, UK.

[3]Key Laboratory of Earth and Planetary Physics, Institute of Geology and Geophysics, Chinese Academy of Sciences, Beijing, China.

[4]College of Earth and Planetary Sciences, University of Chinese Academy of Sciences, Beijing, China.

[5]University of California, Santa Cruz, Santa Cruz, CA, USA.

[6]State Key Laboratory of Lithospheric Evolution, Institute of Geology and Geophysics, Chinese Academy of Sciences, Beijing, China.

[7]Center for Excellence in Comparative Planetology, Chinese Academy of Sciences, Hefei, China.

Corresponding author: Tao Wang (twang0630@gmail.com), Chao Qi (qichao@mail.iggcas.ac.cn)

**Abstract**

Seismic anisotropy, arising from the crystallographic/lattice-preferred orientation of anisotropic minerals or the shape-preferred orientation of melts or cracks, can establish a critical link between Mars's past evolution and its current state. So far, although seismic anisotropy in Mars has been proposed due to different velocities of vertically and horizontally polarized shear waves in the Martian crust, obtained from crustal converted waves, multiples, and surface waves recorded by the InSight seismometer, the evidence is plausible. Notably, the shear wave splitting, which stands out as a straight indicator of seismic anisotropy, has not been reported using marsquake records. In this study, we employ Low-frequency marsquakes detected by the InSight seismometer to reveal shear wave splitting in Mars. We find that the direct $S$ waves of three marsquake recordings (S0173a, S0235b, and S1133c) with high signal-to-noise ratios exhibit the splitting pheonmenon. We rule out the possibility of apparent anisotropy through synthetic tests, affirming the presence of seismic anisotropy in Mars. The delay time (~1.33 s on average) measured from the direct $S$ wave splitting is too large to be solely attributed to the seismic anisotropy in the upper crust (0 – 10 km) beneath the InSight. Thus, seismic anisotropy in the deeper region of Mars is indispensable. Combined with other geophysical evidence near the InSight landing site, the strong seismic anisotropy observed in this study implies the porous crust with aligned cracks being greater than 10 km beneath the InSight and/or the presence of an active mantle plume underneath the Elysium Planitia of Mars.

# 1 Introduction

Seismic anisotropy terms the directional dependence of seismic wave velocities within a medium (Anderson, 1961; Babuska and Cara, 2012), which is result from the crystallographic/lattice-preferred orientation of anisotropic minerals or the shape-preferred orientation of melts or cracks (Almqvist and Mainprice, 2017; Kendall et al., 2005; Long and Becker, 2010; Long and Silver, 2009). Seismic anisotropy is often related to exogenetic process in the shallow crust, such as meteorite impacts (e.g., Beghein et al., 2022; Li et al., 2022), or to endogenic process in the deep interior, such as tectonic stress or mantle flow (e.g., Long & Becker, 2010; Long & Silver, 2009); thus, seismic anisotropy is important for understanding the interior structure and thermal evolution of terrestrial planets. The National Aeronautics and Space Administration's Interior Exploration using Seismic Investigations, Geodesy, and Heat Transport (InSight) mission successfully deployed the Seismic Experiment for Interior Structure (SEIS) instrument on the Elysium Planitia of Mars (Lognonné et al., 2019) (Figure 1A) and obtained continuous seismic waveforms of Mars from February 2019 to December 2022 (Banerdt et al., 2020; Lognonné et al., 2023). Throughout the InSight mission, the Marsquake Service has identified 14 quality A marsquakes (including two meteoroid impact events) with confidently estimated back azimuths and epicentral distances (InSight Marsquake Service, 2023). This provides a unique opportunity to investigate the seismic anisotropy of Mars.

Seismic anisotropy in Mars remains poorly understood. In the upper crust (0 – 8 km) underneath InSight, azimuthal anisotropy was proposed by comparing the seismic velocities of horizontally (*SH*) and vertically (*SV*) polarized shear waves (i.e., shear waves in tangential and radial components), with $V_{SH} < V_{SV}$ (Li et al., 2022). However, the velocities of *SH* and *SV* waves were estimated in a different manner, with the $V_{SH}$ derived from the crustal multiple wave (Li et al., 2022), and the $V_{SV}$ obtained from the receiver function method (Knapmeyer-Endrun et al., 2021). In the top of the crust between 10 – 25 km depth along the seismic ray path from the event S1222a to the InSight station, the presence of radial anisotropy is indicated by the group velocity dispersion analysis of the Love and Rayleigh surface waves from the largest marsquake S1222a, with $V_{SH} > V_{SV}$ (Beghein et al., 2022). The two studies by Li et al., (2022) and Beghein et al., (2022) are not contradictory as the shear wave propagation direction for surface waves (horizontal) is different from that for body waves (vertical). Nevertheless, the seismic anisotropy derived from the inversion of the group velocity dispersions of a single seismic event is elusive, as an isotropic model could also explain the observed surface waves of S1222a (Xu et al., 2023). Last but not least, in the deep crust (> 25 km) or mantle, seismic anisotropy has not been reported, which may provide crucial clues to the dynamics of mantle flow in Mars, as mantle flow has been invoked to explain the formation of the Tharsis Rise and the crustal dichotomy of Mars (e.g., Harder, 2000; Zhong, 2009; Zhong & Zuber, 2001). Therefore, the anisotropic structure in the interior of the Mars, particularly in the deeper region (>10 km), remains unclear but is key to the understanding of the evolutionary history.

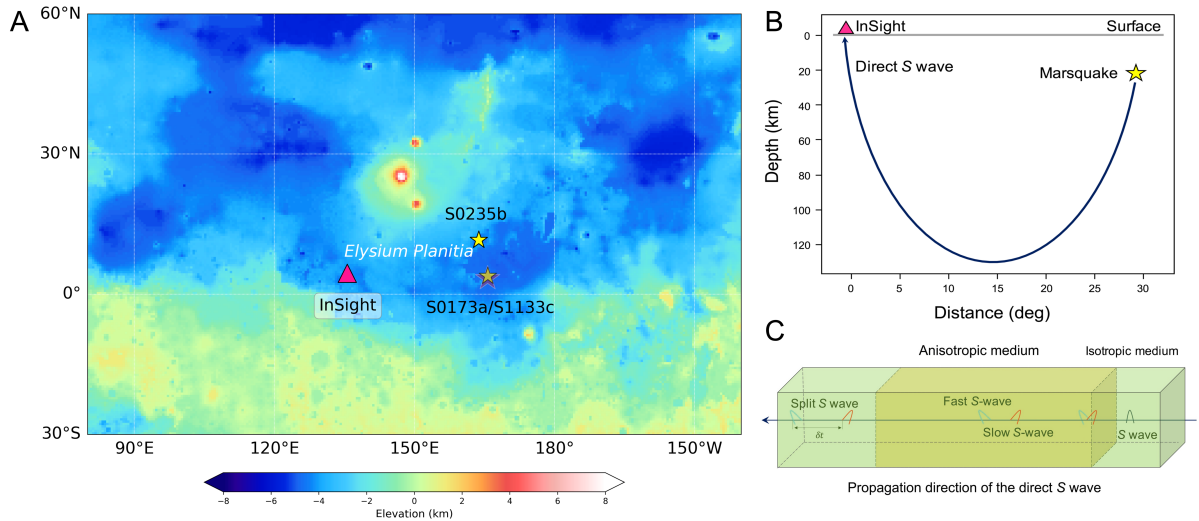

Figure 1. (A) Location of the InSight seismometer (pink triangle) and the selected events (yellow stars). We used the red shadow star to mark event S0173a because the locations of S0173a and S1133c are very close (InSight Marsquake Service, 2023). The map of Mars is from the Mars Orbiter Laser Altimeter (Smith et al., 2001). (B) Cross-section illustration of the direct *S*-wave ray path for the marsuquake with a focal depth of 20 km and an epicentral distance of 30 degrees. The ray path of the direct *S* wave is calcualuted using the velocity model InSight_KKS21GP (Khan et al., 2021; Knapmeyer-Endrun et al., 2021) of Mars and projected in Cartesian coordinates. (C) Schematic of the direct *S* wave splitting. The direct *S* wave (black pulse) propagates from right to left. The direct *S* wave does not split in an isotropic medium (green body), whereas the direct *S* wave splits into the fast shear wave (blue pulse) and the slow shear wave (red pulse) with an accumulated delay time ($\delta t$) after passing through an anisotropic medium (orange body). The polarized directions of the fast and slow shear waves are orthogonal to each other.

The direct *S* wave splitting, as one kind of the shear wave splittings in seismology, is a definitive indicator of seismic anisotropy (Long and Becker, 2010; Long and Silver, 2009). The direct *S* wave refers to the shear wave emanating from the source that propagates directly to the seismic station (Figure 1B). The direct *S* wave is split into two orthogonally polarized components as it propagates through an anisotropic medium (i.e., the direct *S* wave splitting) (Figure 1C). Because the seismic velocities in the two polarization directions are different, fast and slow shear waves are formed with an accumulated delay time (Long and Becker, 2010; Long and Silver, 2009). Meanwhile, the polarization direction of the fast shear wave represents the preferred orientations of mineral fabrics or microstructures within the anisotropic medium (Savage, 1999). On Earth, the direct *S* wave splitting can be measured in local earthquakes (Long and Silver, 2009) and is frequently employed to study seismic anisotropy in the crust and mantle of Earth (e.g., Bi et al., 2020; Huang et al., 2011; Long and Silver, 2009). On Mars, although the direct *S* waves of quality A marsquakes are identifiable, the direct *S* wave splitting has not been detected due to the distortion of the waveform by high-frequency signals (e.g., noises, multiples, or scattered signals) (Li et al., 2022).

In order to reduce the influence of high-frequency signals, this work attempts to conduct the direct *S* wave splitting measurements at low frequencies for quality A marsquakes with high signal-to-noise ratios (SNRs) to determine seismic anisotropy in Mars. In Section 2, we describe the data and methods used in this paper. Then, Section 3 presents the results of the direct *S* wave splitting measurements on the direct *S* waves of marsquakes. In Section 4, we confirm the seismic anisotropy in Mars's interior and delve into its source and implication for the evolution of Mars. Finally, we conclude that beyond the upper crust (0 – 10 km) beneath InSight, seismic anisotropy exists in the deeper region of Mars.

**2 Data and Methods**

In our analysis, we used the 20 samples-per-second seismic data of the three very broadband components (BHU, BUV, and BHW) collected by the SEIS (InSight Mars SEIS Data Service., 2019; Lognonné et al., 2019). We first removed glitches following the synthetic template technique of Scholz et al., (2020), then deconvolved the instrumental response from the raw data to obtain the velocity records. Next, we rotated the seismic waveforms from the U (BHU)-V (BHV)-W (BHW) system to the geographical azimuth system (north: BHN, east: BHE, and vertical: BHZ) according to the dip and azimuth angles of the InSight's seismic sensors (Ceylan et al., 2021).

We employed Low-frequency seismic events with a location quality of A in the marsquake catalog (InSight Marsquake Service, 2023). The InSight Marsquake Service classifies marsquakes into Low-frequency and High-frequency families, with the former dominated by long-period (including low-frequency and broadband) signals and the latter by high-frequency signals (Clinton et al., 2021; Giardini et al., 2020). The Low-frequency family is interpreted as marsquakes occurring in the deep crust or mantle, generally with an identified direct *S* wave and little energy trapped in the crust (Giardini et al., 2020), while the High-frequency family is attributed to marsquakes occurring in the shallow crust with an unidentified direct *S* wave and most of the energy trapped in the crust, leading to many scattered waves in the seismic waveforms (Menina et al., 2021; Van Driel et al., 2021). Therefore, to use the direct *S* wave and minimize the effect of source-side scattering, we chose seismic events from the Low-frequency family. We did not use the two meteorite impact events (S1000a and S1094b), which belong to the quality A marsquakes, because the epicentral distance of S1000a was too large (~128°) to produce a direct *S* wave, and the SNR of the direct *S* wave of S1094b is low (Posiolova et al., 2022).

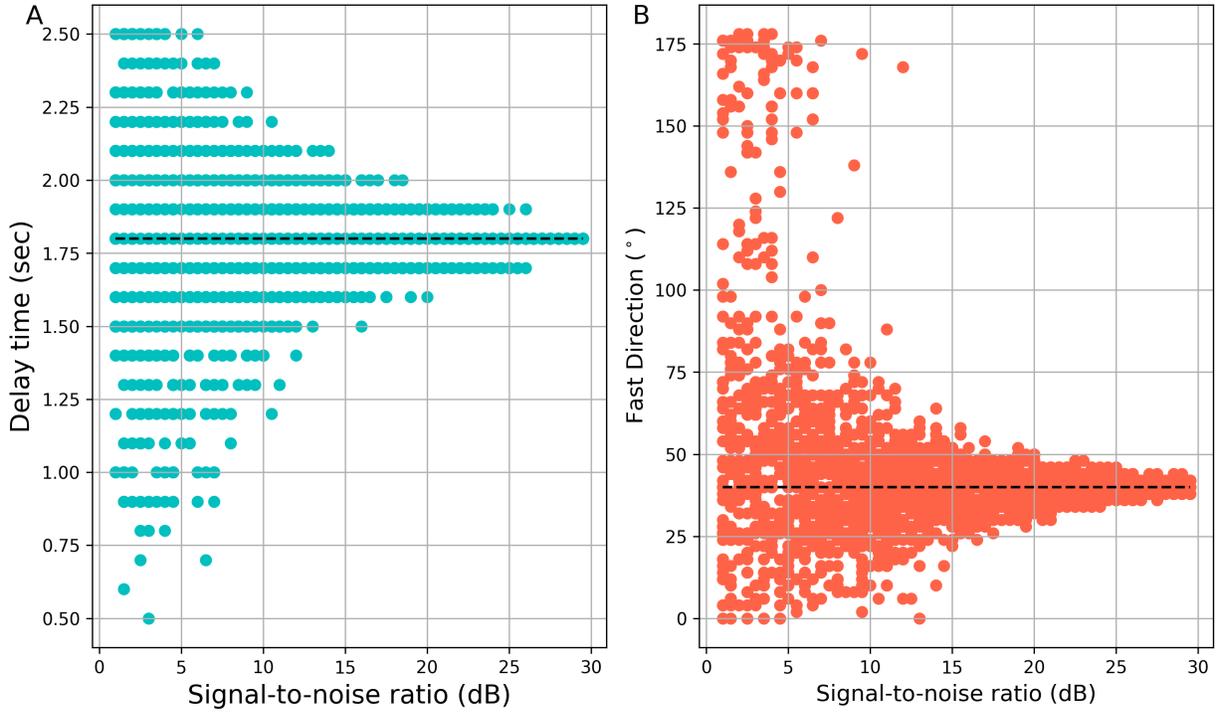

Figure 2. Effect of signal-to-noise ratio on the shear wave splitting parameters. The black horizontal dashed line in each panel marks the preset splitting parameters, with a delay time of 1.8 s and a fast direction of 40°. For each signal-to-noise ratio, we repeated it 50 times to generate random noise and then measured the splitting parameters through the eigenvalue method (Silver and Chan, 1991). The cyan and orange dots in panels (A) and (B) represent the measured delay time and fast direction for each round, respectively.

We calculated the SNR of the direct *S*-wave of a seismic event in both the frequency and time domains using Equation 1:

$$SNR = 10 \cdot log_{10} \left(\frac{A_S}{A_N}\right)^2. \tag{1}$$

In the frequency domain, $A_S$ and $A_N$ are the root mean squares of the noise and signal amplitude spectra (in 20-s time windows before and after the direct *S* wave), respectively. In the time domain, $A_S$ and $A_N$ are the root mean squares of the noise and signal waveforms, respectively. The MQS (InSight Marsquake Service, 2023) provides all of the direct *S*-wave arrival times of Low-frequency marsquakes. We utilized the frequency band between 0.1 Hz and 0.3 Hz because, at frequencies other than 0.1 – 0.3 Hz, the direct *S*-wave signal is difficult to distinguish from noises, as shown in the amplitude spectra (Figures S1 and S2 in the Supporting Information). We selected only high SNR events (SNR>10.0 dB for both horizontal components and both frequency and time domains) (Figures S1, S2, and S3 in the Supporting Information) because the noise can significantly affect the shear wave splitting measurements (e.g., Figure 2). Finally, we retained three events, including S0173a, S0235b, and S1133c (Figure 3), among which, S0235b has the highest SNR in both the frequency and time domains (>20.0 dB).

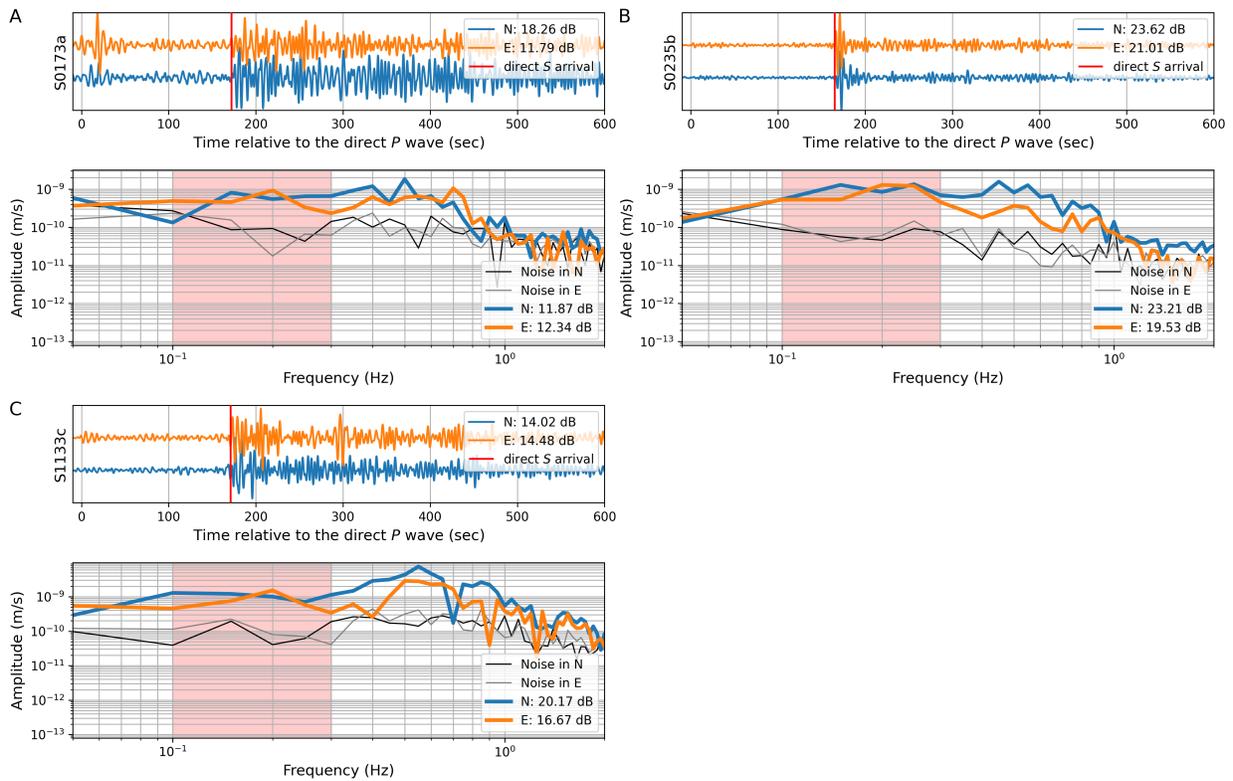

Figure 3. (A - C) The seismograms (top) and the corresponding amplitude spectra (0.05 – 2 Hz) (bottom) of selected events. In the seismograms panel, the blue and orange solid lines represent the northern (N) and eastern (E) seismic waveforms filtered between 0.1 – 0.3 Hz, respectively. The X-axis is the time relative to the direct *P* wave, and the red vertical line marks the direct *S* wave arrival provided by the InSight Marsquake Service (2023). In the amplitude spectrum panel, the colored solid lines denote the amplitude spectra of the signal windows (blue: northern component, orange: eastern component). The black (N component) and gray (E component) solid lines represent the amplitude spectra of noise windows. The red shadow indicates the frequency range (0.1 – 0.3 Hz) used for the direct *S* wave splitting analysis. The noise and signal windows used to calculate the amplitude spectra are the 20-s time windows before and after the direct *S* wave, respectively. The signal-to-noise ratio of the direct *S* wave of each event in the time domain and frequency domain is labeled on the right.

We conducted the direct $S$ wave splitting measurements using the eigenvalue method (Silver and Chan, 1991) to obtain the splitting parameters (fast direction: $\phi$, and delay time: $\delta t$) since the initial polarization of the direct $S$ wave of the marsquake is unknown. In the grid search, $\phi$ ranges from 0° to 180° with an increment of 1°. To avoid the cycle skip, the search range for $\delta t$ is 0 – 2.5 s (half a period of the direct $S$ wave) with an increment of 0.05 s (corresponds to the sampling rate) because the source radiation pattern or the dipping interface might cause a polarity reversal in the two horizontal components. As the choice of the time window of the direct $S$ wave can affect the measurement of the splitting parameters (Teanby, 2004), we measured the splitting parameters in 25 different time windows and then carried out cluster analysis (e.g., Figure 4). The start of the time window varies from -5 s to 0 s before the direct $S$ wave with an interval of 1 s, and the end is from 8 s to 10 s after the direct $S$ wave with an interval of 0.5 s (e.g., Figure 4A). Ending windows cannot be too late to avoid interference from crustal multiples (Li et al., 2022). To ascertain the presence of shear wave splitting, we applied three criteria: convergence degree, particle trajectory, and cluster analysis. Specifically, we considered the shear wave to be split only if the measurement map converges to an extremum (e.g., Figure 4B), the particle motion trajectory of the waveform exhibits elliptical before correction and becomes linear after correction (e.g., Figure 4C), and the cluster analysis is convergent (e.g., Figure 4E).

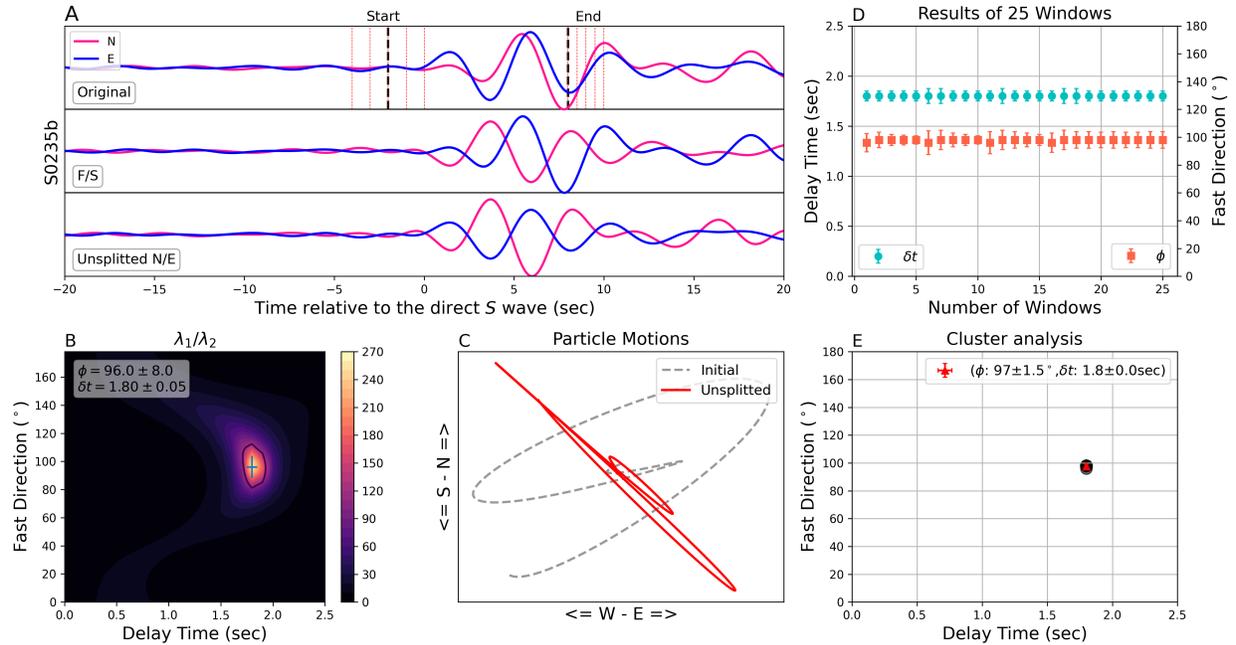

Figure 4. Direct *S* wave splitting measurements from the marsquake recording of S0235b. (A) Top: The direct *S*-wave seismograms (0.1 – 0.3 Hz) in two horizontal components of event S0235b. The pink and blue solid lines represent the northern (N) and eastern (E) components, respectively. The X-axis is the time relative to the arrival of the direct *S* wave. The red vertical dashed lines mark the start and end offsets of the time windows used for the direct *S* wave splitting measurements. The black vertical dashed lines mark the time window, within which the waveforms are used to show the examples of the measurement map and particle motion. Middle: Fast (F) (in pink) and slow (S) (in blue) transverse waveforms after rotating the two horizontal components into the fast and slow directions. Bottom: Unsplitted N (pink) and E (blue) components after correcting the fast direction and delay time. (B) Measurement map of the eigenvalue method. The black solid lines are the isolines at the 95% confidence level estimated using the F-test. The blue cross indicates the searched optimal splitting parameter, with the length representing the one sigma error. (C) Particle motions before (gray) and after (red) correction. (D) Splitting parameters are measured over 25 time windows. The cyan dots and orange squares denote the delay time ($\delta t$) and the fast direction ($\phi$), respectively. The error bar corresponds to the one sigma error. (E) Cluster analysis for the 25 optimal splitting parameters from (D). The black-filled circles represent the splitting parameters measured in each time window. The red-filled triangle indicates the solution of the cluster analysis, with the error bar corresponding to the two standard deviations.

## 3 Results

The direct *S* waves from all the three marsquakes (S0173a, S0235b, and S1133c) exhibit the splitting phenomenon, as shown in Figure 4, 5, and 6. The fast directions obtained from the direct *S* wave splitting measurement in the three events are generally consistent (68° – 96°), with differences of less than 30 degrees. Among the three events, S0235b, with the highest SNR, has the largest delay time of $1.8^{\pm0.05}$ s, followed by S0173a with $1.3^{\pm0.35}$ s, and S1133c with the smallest delay time of $0.9^{\pm0.28}$ s; thus, the average delay time is about 1.33 s.

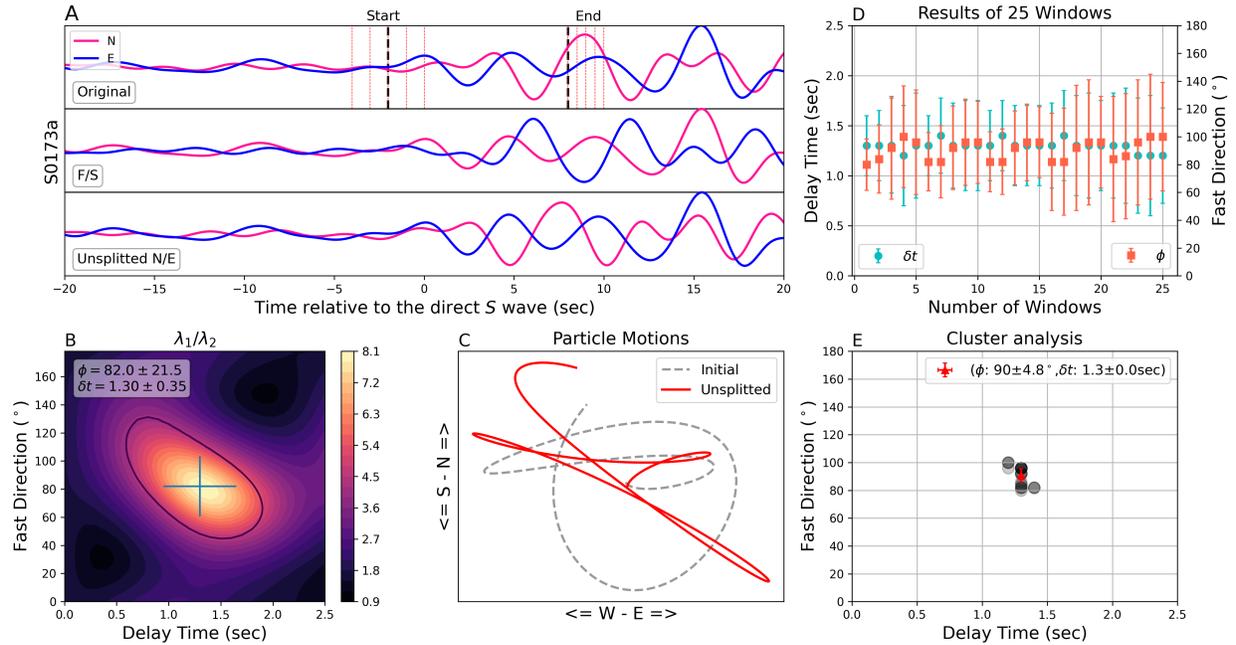

Figure 5. Direct *S* wave splitting measurements for the marsquake recording of S0173a. The caption of each panel is the same as that in Figure 4.

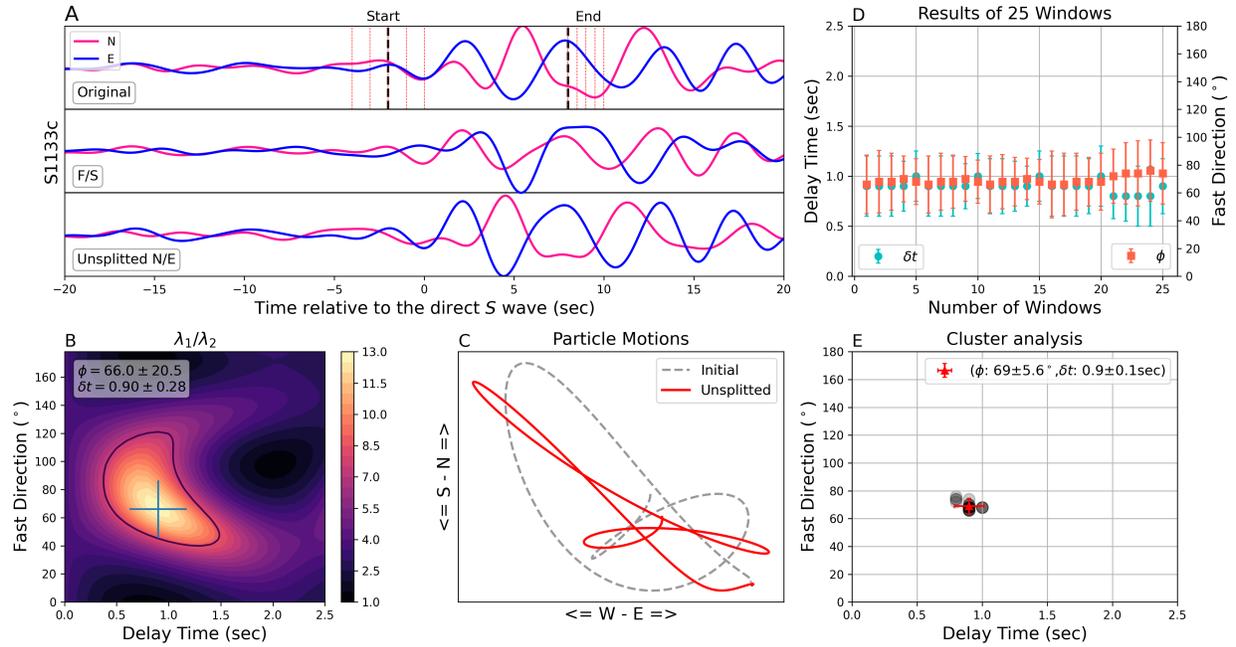

Figure 6. Direct *S* wave splitting measurements for the marsquake recording of S1133c. The caption of each panel is the same as that in Figure 4.

The measurement uncertainties of the fast direction and the delay time of S0235b are the smallest compared to those of S0173a and S1133c (Table 1), which can be explained by the following two reasons. Firstly, the SNR of S0235b is the highest among the three events (~22.3 dB, while ~15 dB for the other two events, as shown in Figure 3). Synthetic tests demonstrate that the uncertainties of the splitting parameters are small ($\Delta\phi < 10°$, $\Delta\delta t < 0.05$ s) when the SNR exceeds 20.0 dB (see Figure 2). Secondly, in the case of S0235b, the amplitude of the direct *S* wave surpasses that of the subsequent coda waves (Figure 3B). In contrast, for S0173a and S1133c, the amplitudes of the direct *S* wave and the subsequent coda wave are comparable (Figure 3A and 3C). This observed amplitude difference is also likely to contribute to larger measurement uncertainties in S0173a and S1133c. Consequently, the result of S0235b emerges as the most compelling evidence supporting the direct *S* wave splitting phenomenon in the marsquake recording, while those of S0173a and S1133c serve as reference.

Table 1. Summary of the back azimuth, epicentral distance, and slowness information of S0173a, S0235b and S1133c (InSight Marsquake Service, 2023) and their corresponding splitting parameters. The superscript is the one sigma error estimated from the measurement map of the eigenvalue method (e.g., Figure 4B). We also calculated the average signal-to-noise ratio of their direct *S*-wave in the time domain of the two horizontal components.

| Event | Back azimuth (degree) | Epicentral distance (degree) | Slowness (s/km) | Fast direction (degree) | Delay time (sec) | Signal-to-noise ratio (dB) |
|---|---|---|---|---|---|---|
| S0173a | 86.6 | 30.0 | 0.22 | $89^{\pm 21}$ | $1.3^{\pm 0.35}$ | ~15.0 |
| S0235b | 77.0 | 28.8 | 0.22 | $96^{\pm 8}$ | $1.8^{\pm 0.05}$ | ~22.3 |
| S1133c | 86.4 | 30.0 | 0.22 | $68^{\pm 20}$ | $0.9^{\pm 0.28}$ | ~14.2 |

The core-transiting *SKS* wave is the most widely utilized phase for detecting seismic anisotropy in Earth, as the *SKS* wave splitting only reflects receiver-side anisotropy between the surface and the core-mantle boundary (Long and Silver, 2009). However, in the case of Mars, due to the very low SNR of the previously identified *SKS* phases in marsquake recordings (Irving et al., 2023) (Figure S4 in the Supporting Information), we cannot further verify the shear wave splitting using the *SKS* phase in the marsquake recording. In addition, the waveform distortion caused by the high-frequency signals (see the amplitude spectra in Figure 3) hinders our attempt to obtain reliable splitting parameters in a high-frequency band of 0.1 – 0.8 Hz (Figure S5 in the Supporting Information).

4 Discussion

4.1 Confirmation of the anisotropic structure

Shear wave splitting can result from the interaction of different seismic phases in an isotropic medium (e.g., Parisi et al., 2018) (i.e., apparent anisotropy) or from the transverse seismic wave propagating through an anisotropic medium (Long and Silver, 2009). In order to rule out the possibility of apparent anisotropy caused by multiples, we restricted the end of the time window (up to 10 s after the direct *S* wave) in the direct *S* wave splitting measurement to avoid the influence of crustal multiples originating from the upper crust near the InSight (Li et al., 2022). Furthermore, we used the program AxiSEM3D (Leng et al., 2016) to calculate synthetic seismic waveforms in isotropic media and measured the direct *S* wave splitting in the same way as we measured that for real data. We simulated the seismic waveforms of S0235b because of the highest SNR of the direct *S* wave and the minimal measurement uncertainties of the direct *S* wave splitting for S0235b, compared to those for S0173a and S1133c.

The input parameters for the simulation are set based on previous researches of Mars (Drilleau et al., 2022; Durán et al., 2022; InSight Marsquake Service, 2023; Jacob et al., 2022; Joshi et al., 2023; Knapmeyer-Endrun et al., 2021; Shi et al., 2023a). We utilized two different Martian velocity models ([Table S1 in the Supporting Information](#)): one featuring a 3-layer crustal structure (Durán et al., 2022; Joshi et al., 2023; Knapmeyer-Endrun et al., 2021) from the velocity model InSight_KKS21GP (Khan et al., 2021; Knapmeyer-Endrun et al., 2021) of Mars and the other with a slight difference, incorporating an additional low-velocity layer of 2 km on top (Shi, et al., 2023a) (i.e., 4-layer crustal structure). We set the source moment tensor of S0235b (strike: $75°$, dip: $72°$, rake: $-110°$) in the case of 21-km depth with reference to the analysis of seismic sources (Jacob et al., 2022) and depth phases (Drilleau et al., 2022) of marsquakes. The epicentral location of S0235b is provided by the InSight Marsquake Service (2023), as shown in Figure 1A. Given the unknown real source time function for S0235b, we employed the Gaussian source time function for this event. Besides, the maximum frequency of the simulation is set to be 0.8 Hz to capture as many multiples as possible. We filtered the synthetic waveforms with a 0.1 – 0.3 Hz Butterworth bandpass filter and then measured the splitting parameters in the same way as the real data.

It is unlikely that the direct *S* wave splitting observed marsquakes arises from the apparent anisotropy of the multiples. Figure S6 (in the Supporting Information) demonstrates that in the isotropic media, three criteria mentioned in Section 2 can help to rule out the possibility of apparent anisotropy, i.e., no notable convergence in the measurement map of the eigenvalue method, no significant variation in the particle motion before and after the correction, and no convergence in the cluster analysis. In order to eliminate the potential influence of source information uncertainty, we conducted additional tests using the Ricker source time function and a different source moment tensor of S0235b (strike: $55°$, dip: $15°$, rake: $-134°$) obtained by Brinkman et al., (2021). The results (Figures S7 and S8 in the Supporting Information) again demonstrate that in the isotropic case, the synthetic direct *S* wave shows no splitting. In contrast, we can observe the similar direct S-wave splitting phenomenon when we add the *S*-wave anisotropy into the synthetic model (e.g., Figure S9 in the Supporting Information). Therefore, it is seismic anisotropy in Mars's interior that leads to the direct *S* wave splitting observed in the marsquake recordings of S1073a, S0235b, and S1133c.

4.2 Insufficient anisotropy in the upper crust

The measured fast directions ($68°$ – $96°$) are consistent with the direction of the azimuthal anisotropy in the upper crust obtained by Li et al. (2022), while the measured delay time (~1.33 s on average) is so large that the source of the anisotropy needs to be discussed. In this section, we carefully compared our results with the seismic anisotropy in the upper crust obtained by Li et al. (2022), aiming to determine whether seismic anisotropy exists in Mars's deeper interior beyond the upper crust (0 – 10 km) below InSight.

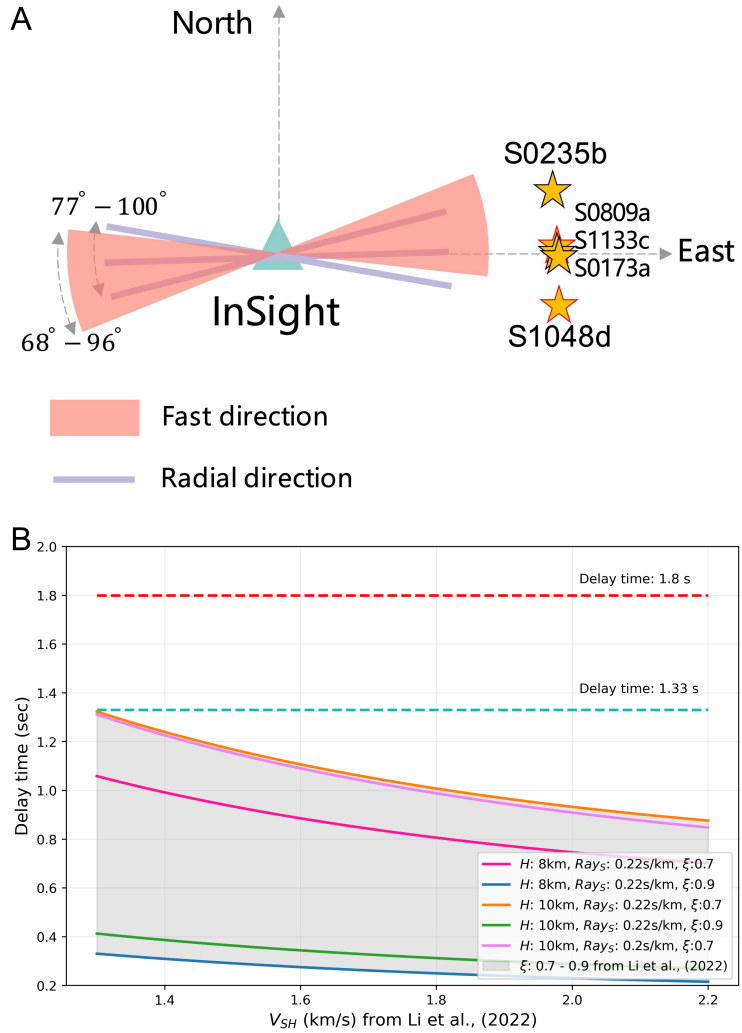

Figure 7. (A) Diagram of the direction of the azimuthal anisotropy. The pink shaded area indicates the fast direction obtained by the direct *S* wave splitting (Table 1). Orange stars with black borders are the events used in this study. The purple solid lines denote the radial directions according to the marsquake events (all the five stars) used by Li et al. (2022). The cyan triangle symbolizes the InSight lander, with the two directional arrows marking the north and east orientations. (B) The variation of the delay time with the velocity of the horizontally polarized shear wave ($V_{SH}$). The delay time mainly depends on the value of the anisotropy coefficient, $\xi = \left(\frac{V_{SH}}{V_{SV}}\right)^2$. The anisotropy coefficient in the upper crust beneath InSight is estimated between 0.7 – 0.9 (the gray shade) by Li et al. (2022), which is insufficient to account for the measured delay time of the direct *S* wave splitting (~1.33 s on average and 1.8 s for event S0235b). Different coloured curves represent the dependence of delay time on $V_{SH}$ across different upper crustal thicknesses ($H$), *S*-wave ray parameters ($Ray_S$), and anisotropy coefficient.

The directions of the azimuthal anisotropy obtained by Li et al. (2022) and this study are in general agreement. Through a comparison of the *SV* wave (shear wave in the radial direction) velocity ($V_{SV}$) derived from receiver function analysis (Knapmeyer-Endrun et al., 2021) and the *SH* wave (shear wave in the tangential direction) velocity ($V_{SH}$) inferred from the crustal multiple wave *SsSs* identified in the tangential component, Li et al. (2022) reach the conclusion that $V_{SV} > V_{SH}$. Such seismic anisotropy reflects azimuthal anisotropy in the radial and tangential directions with the radial shear wave being faster. The radial directions range from 77° - 100° according to the back azimuths of events used by Li et al. (2022) (The purple solid lines in Figure 7A). In the context of this study, if the predominant origin of seismic anisotropy is in the crust, then the direct *S* wave splitting phenomenon is also indicative of azimuthal anisotropy. The fast direction, measured in the direct *S* wave splitting, indicates the direction of the fast shear wave. The determined fast directions are between 68° - 96° (Table 1) (The pink shaded area in Figure 7A), overlapping with the radial directions. Consequently, if the crustal seismic anisotropy in Mars is the primary source responsible for the direct *S* wave splitting, the direction of our obtained azimuthal anisotropy aligns with that reported by Li et al. (2022).

However, the delay time arising from seismic anisotropy in the upper crust inferred by Li et al. (2022) cannot solely account for the delay time measured from the direct *S*-wave splitting in marsquake recordings. A direct comparison between the delay time derived from the upper crust and the delay time measured in the direct *S* wave splitting is feasible, as the radial directions of the events employed by Li et al. (2022) and the directions of the fast shear wave obtained in this study are basically consistent (Figure 7A). More specifically, the anisotropy coefficient $\xi = \left(\frac{V_{SH}}{V_{SV}}\right)^2$ in the upper crust beneath InSight falls within the range of 0.7 to 0.9 (Li et al., 2022). The uncertainty of $\xi$ stems from the exclusive reliance on inferring the upper crustal thickness (*H*) and $V_{SH}$ through the *SsSs* wave in the upper crust, as a substantial trade-off exists between thickness and wave velocity. The upper crustal thickness is estimated between 8 km and 10 km, and $V_{SH}$ is 1.3 – 2.2 km/s (Li et al., 2022). Adpoting the same ray parameter ($Ray_S$) of 0.22 s/km as the real events (Table 1) and the ray parameter of 0.2 s/km for comparison, one can calcualute the delay time ($\delta t$) induced by seismic anisotropy in the upper crust using Equation 2:

$$\delta t = \frac{V_{SV} - V_{SH}}{V_{SV} \times V_{SH}} \times \frac{H}{\sqrt{1 - Ray_S^2 \times \left(\frac{V_{SV} - V_{SH}}{2}\right)^2}}, \tag{2}$$

where $V_{SV} = \frac{V_{SH}}{\sqrt{\xi}}$. Figure 7B reveals that, regardless of whether the upper crustal thickness beneath InSight is 8 km or 10 km, the delay time induced from the upper crust is insufficient to account for the most reliable delay time of 1.8 s (red horizontal dashed line in Figure 7B) obtained from the splitting parameter in S0235b, and even in the case of the average delay time of 1.33 s (cyan horizontal dashed line in Figure 7B) from the splitting parameters in S0173a, S0235b, and S1133c (Table 1). Moreover, Figure 7B shows that $\xi$ has a greater effect on the delay time than the thickness of the upper crust and the *S*-wave ray parameter. The delay time increases with decreasing $\xi$ (i.e., increasing anisotropy). If seismic anisotropy primarily resides in the upper crust, the magnitude of seismic anisotropy required to account for the direct *S* wave splitting would be greater than at least ~20% ($\xi$<0.7) (Figure 7B), a threshold that contrasts with the magnitude of seismic anisotropy within the upper crust (<20%) ($\xi$>0.7) (Li et al., 2022). Therefore, the magnitude of seismic anisotropy in the upper crust beneath InSight (Li et al., 2022) cannot fully interpret the splitting delay time.

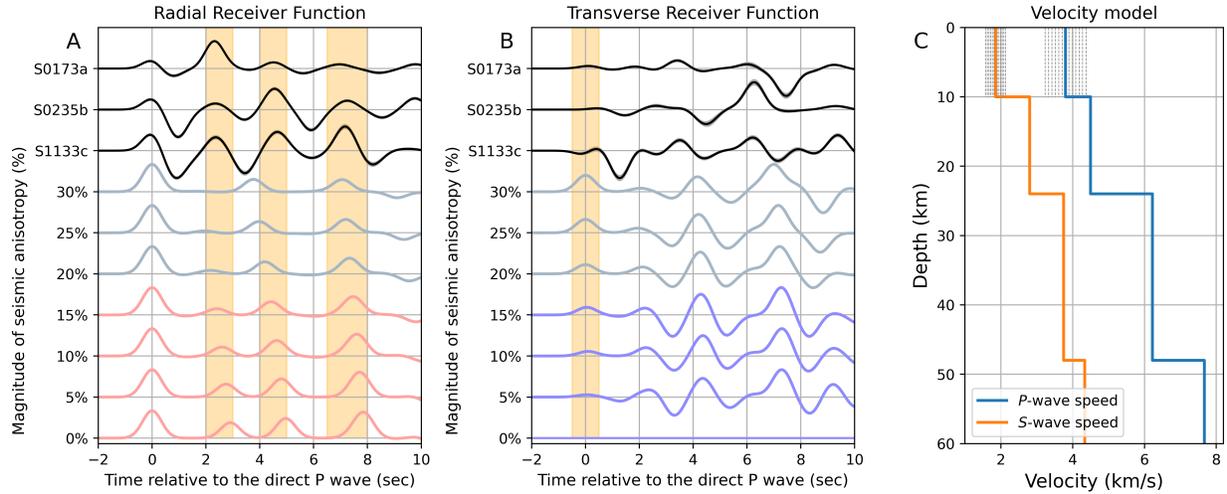

Figure 8. The observed and synthetic radial (A) and transverse (B) receiver functions. In panels (A) and (B), the black solid lines with gray shades represent the observed receiver functions with two standard deviations taken from Shi et al, (2023a). The three orange shades in (A) indicate the three converted phases observed in other studies with different approaches (Dai & Sun, 2023; Durán et al., 2022; Joshi et al., 2023; Kim et al., 2021; Knapmeyer-Endrun et al., 2021; Lognonné et al., 2020; Shi et al., 2023b). The orange shade in (B) denotes the direct *P* wave. The red (in panel A) and purple (in panel B) solid lines represent the synthetic receiver functions for media with anisotropies less than 20%, while the gray solid lines are for those greater than 20%. The synthetic receiver functions are calculated using the procedure Telewavesim (Audet et al., 2019). (C) Velocity model used to calculate the receiver functions. This is model is from the three-layer crustal model of Knapmeyer-Endrun et al., (2021). The gray vertical dashed lines mark the different magnitudes of the anisotropies added in the first layer.

This contradiction can be further verified using the *P*-wave receiver function (PRF) (Ligorria and Ammon, 1999), a method that can also be used to measure seismic anisotropy (e.g., Liu & Niu, 2012; Nagaya et al., 2008). We acquired the real radial and transverse PRFs of S0173a, S0235b, and S1133c from our previous PRF study (Shi et al., 2023a) and compared them with the synthetic PRFs calculated from various models where the anisotropy is added in the upper crust between 0 km and 10 km. The real radial PRFs (black lines in Figure 8A) reveal four prominent positive phases at 0 s, ~2.4 s, ~4.5 s, and ~7.2 s. The first corresponds to the direct *P* wave, and the latter three are identified as crustal *P*-to-*S* waves (Dai & Sun, 2023; Durán et al., 2022; Joshi et al., 2023; Kim et al., 2021; Knapmeyer-Endrun et al., 2021; Lognonné et al., 2020; Shi et al., 2023b). Meanwhile, the real transverse PRFs exhibit weak energy at 0 s (black lines in Figure 8B). Theoretical tests demonstrate that when the anisotropy of the upper crust is less than 20%, the synthetic radial and tangential PRFs (coloured lines in Figure 8A and 8B) align well with the characteristics of real PRFs. However, when the upper crust anisotropy exceeds 20%, the synthetic radial PRFs fail to capture the two positive phases at 2.4 s and 4.5 s, while the synthetic transverse PRFs display strong direct $P$ waves at 0 s, which are also inconsistent with the real ones. Consequently, analysis of the direct $S$ wave splitting, together with the analysis of the PRFs, reveals that seismic anisotropy exists elsewhere in the deeper region of Mars rather than solely in the upper crust between 0 km and 10 km beneath InSight. However, we cannot distinguish the specific source region of the anisotropy because the direct $S$ wave splitting measured in this study represents the integral of the direct $S$-wave splitting along the entire seismic path from source to station, which lacks the depth resolution.

4.3 Two possible mechanisms of the deep anisotropy

The strong seismic anisotropy needed to account for the direct $S$ wave splitting requires the contribution of seismic anisotropy in the deeper region (> 10 km), which can result from two possible mechanisms. One mechanism is the aligned cracks in the deeper crust beneath InSight. The aligned cracks can result in a tilted and/or horizontal transverse isotropic medium (TTI and HTI), manifesting azimuthal seismic anisotropy (Huang et al., 2015; Shapiro, 2017). All of the direct $S$ waves used in this study (Table 1) have near-vertical incidence in the crust beneath the InSight (about 20 degrees off the vertical axis). Thus, if the direct $S$ wave splitting occurs mainly in the crust, it likely indicates azimuthal anisotropy caused by the aligned cracks. Another mechanism is the crystallographic/lattice-preferred orientation due to the deformation of materials underneath Elysium Planitia. The deformation of materials caused by mantle flow can lead to the crystallographic/lattice-preferred orientation of minerals, thereby producing seismic anisotropy (e.g., Long & Becker, 2010; Long & Silver, 2009). As the direct *S*-wave paths of S0173a, S0235b, and S1133c all pass through the mantle underneath Elysium Planitia, the direct *S*-wave splittings observed in the three marsquakes signify the deformation of materials there.

The possible presences of aligned cracks in the deeper crust suggests that the thickness of the porous crust beneath the InSight is greater than 10 km (Figure 9). Two pieces of evidence support the presence of aligned cracks. Firstly, the InSight lander is located in Elysium Planitia, and the nearby area of the InSight shows near north-south trending wrinkle ridges (Golombek et al., 2020, 2018), suggesting the presence of regional east-west compressive stress (Li et al., 2022). Geodynamic simulations also show the presence of east-west stress at 20 km depth beneath InSight (Broquet and Andrews-Hanna, 2022), which is compatible with the fast directions (68° - 96°) obtained the direct *S* wave splitting (Figure 7A). Secondly, cracks can extend into the deeper crust (> 10 km). The average porosity of the bulk crust of Mars is 10 - 23 % (Goossens et al., 2017), and the porosity of the upper crust (< 10 km) beneath the InSight has been estimated to be greater than ~20% using various seismic techniques (Dai and Sun, 2023; Li et al., 2023); thus, it is conceivable that the deeper crust (> 10 km) still contains cracks. This is supported by the estimate that the base of the porous materials (12 - 23 km) below the InSight is between 12 – 23 km (Wieczorek et al., 2022). As a result, due to the oriented stress and deep-seated cracks, the deep crust is created as an anisotropic medium with the fast shear wave in the east-west direction. In this case, the base of the pore closure is deeper than the previously interpreted depth of 8 – 11 km (Gyalay et al., 2020), indicating a reduced maximum heat flux value (< 60 mW m$^{-2}$) at the InSight site experienced after the pore formation. Meanwhile, such a porous crust (>10 km thickness) can guide our search for water in the deeper crust of Mars (Clifford et al., 2010; Kilburn et al., 2022), since most of the initial water on Mars (30 – 99%) is estimated to have been sequestered by the crustal hydration (Scheller et al., 2021), while rock physics models demonstrate that the shear wave velocity in the upper crust beneath the InSight is too low to suggest a cryosphere there (Manga and Wright, 2021).

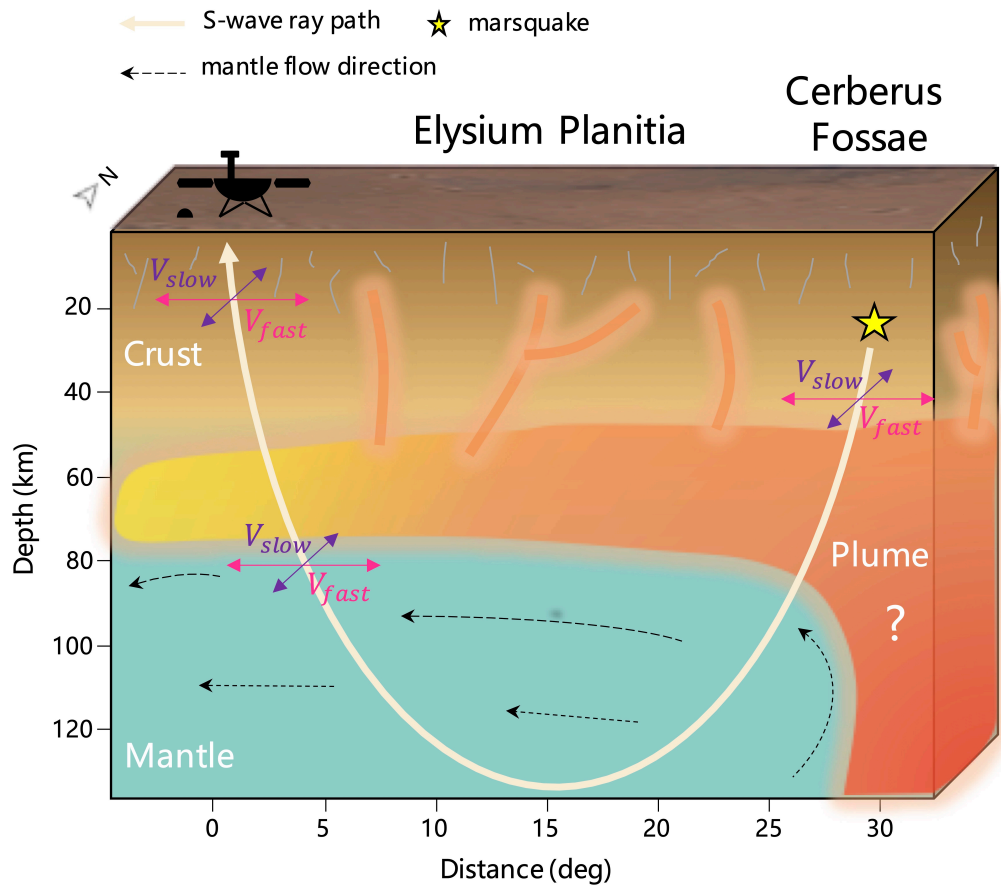

Figure 9. A possible model of the interior structure of Mars for accounting for the direct *S* wave splitting in the marsquake recording. This model incorporates the potential mantle plume beneath Elysium Planitia (Broquet and Andrews-Hanna, 2022). The flesh solid arrowed line represents the ray path of the direct *S* wave from the marsquake (yellow star) to the InSight. Crossed bidirectional arrows (fast shear wave in pink and slow shear wave in purple) mark possible regions of seismic anisotropy in Mars, some originating from the deeper crust (> 10 km) beneath InSight due to the aligned cracks, and others from the Elysium Planitia due to the mante plume.

The deformation of materials underneath Elysium Planitia of Mars suggests possible existence of a mantle plume there (Figure 9). Combining evidence from gravity, topography, volcanism, regional stress, seismicity, and simulations, Broquet & Andrews-Hanna (2022) suggested an active mantle plume beneath Elysium Planitia, with the plume head below the Cerberus Fossae.

Such a mantle plume could create seismic anisotropy in two ways. One way is that due to the stress filed and presence of melt induced by the upwelling mantle plume, an anisotropic medium with a fast shear wave in the east-west direction is formed in the deep crust or mantle near the plume head. Since events S0173a, S0235b, and S1133c are located in the Cerberus Fossae (InSight Marsquake Service, 2023; Wang et al., 2023), the direct *S* waves of the three marsquakes sample such an anisotropic region and exhibit the splitting phenomenon. Another way involves the spreading of mantle plumes, resulting in mantle flow, thus forming seismic anisotropy. However, the limited number and inadequate azimuthal coverage of high-SNR marsquakes restrict our ability to further constrain the mantle plume underneath Elysium Planitia, which may be conducted in combination with other techniques in the future.

Note that the exact location of the strong seismic anisotropy within Mars remains unsettled in the context of this study. We emphasise that the magnitude of seismic anisotropy in the upper crust (< 10 km) alone inadequately accounts for the direct-S wave splitting; hence, seismic anisotropy in the deeper crust, mantle, or a combination thereof is necessary.

**5 Conclusions**

In conclusion, this study presents the first observation of the direct *S* wave splitting in marsquake recordings and provides compelling evidence for seismic anisotropy underneath Elysium Planitia. Our analysis of the direct *S* waves from three high-SNR marsquakes, namely S0173a, S0235b, and S1133c, originating from the Cerberus Fossae, consistently reveals the presence of the splitting phenomenon. Among these events, the results from S0235b are considered the most robust, given its minimal measurement errors and highest SNR. The determined fast directions measured from the direct *S* wave splittings of these three events are generally west-east oriented, ranging from $68°$ to $96°$. Notably, the substantial delay time observed, with an average of approximately 1.33 s (with S0235b registering at 1.8 s), cannot be solely attributed to seismic anisotropy in the upper crust (0 – 10 km) beneath InSight. This suggests the existence of seismic anisotropy in the deeper region of Mars.

The intriguing finding suggests the presence of a thick crust (>10 km) with aligned cracks beneath InSight, and potentially provides evidence for the mantle plume beneath Elysium Planitia of Mars. This yields fresh insights into the evolutionary processes shaping the red planet.

**Declaration of competing interest**

The authors declare that they have no known competing financial interests or personal relationships that could have appeared to influence the work reported in this paper.

**Acknowledgments**

This research is funded by the National Natural Science Foundation of China (Grants 42125303, 42174214) and the Fundamental Research Funds for the Central Universities (Grant 020614380122). Ling Chen is supported by the National Natural Science Foundation of China (42288201). Chao Qi is supported by the Key Research Program of the Institute of Geology and Geophysics, CAS, grant IGGCAS-201905. The authors acknowledge NASA, CNES, their partner agencies and institutions (UKSA, SSO, DLR, JPL, IPGP-CNRS, ETHZ, IC, MPS-MPG) and the flight operations team at JPL, SISMOC, MSDS, IRIS-DMC and PDS for providing SEED SEIS data. Part of the direct $S$ wave splitting measurement is conducted using the procedure SplitWavePy developed by Jack Walpole (https://splitwavepy.readthedocs.io/en/latest/).

**Open Research**

Seismic data of InSight SEIS are provided by InSight Mars SEIS Data Service., (2019). Figures are plotted by Matplotlib (Hunter, 2007).